\documentclass[aps,superscriptaddress,eqsecnum,nofootinbib,showpacs,preprintnumbers]{revtex4-2}
\usepackage[utf8]{inputenc}
\usepackage{amsmath}
\usepackage{amssymb}
\usepackage{amsfonts,amsthm,bm}
\usepackage{color,xcolor}
\usepackage{comment}
\usepackage{soul}

\usepackage{graphicx,epsfig}
\usepackage{float}
\usepackage{subcaption}
\usepackage{tikz}
\newcommand{\be}{\begin{eqnarray}}
	\newcommand{\ee}{\end{eqnarray}}
\newcommand{\bea}{\begin{eqnarray}}
	
	\newcommand{\eea}{\end{eqnarray}}

\newcommand*\diff{\mathop{}\!\mathrm{d}}

\def\del{\partial}

\def\Ref{\ref}

\definecolor{azure(colorwheel)}{rgb}{0.0, 0.5, 1.0}
\definecolor{DarkViolet}{RGB}{148,0,211}
\definecolor{myDarkBlue}{rgb}{0,0.1,0.7}
\definecolor{DarkBlue}{RGB}{0,0,153}
\definecolor{amber}{rgb}{1.0, 0.49, 0.0}
\definecolor{amaranth}{rgb}{0.9, 0.17, 0.31}
\definecolor{nicered}{rgb}{0.7,0.1,0.1}
\definecolor{brown}{rgb}{0.5,0.1,0.1}
\definecolor{nicegreen}{rgb}{0.0,0.3,0.0}
\definecolor{tealgreen}{rgb}{0.0, 0.51, 0.5}

\definecolor{tclr}{RGB}{148,0,211}

\usepackage{hyperref}
\hypersetup{colorlinks,bookmarksopen,
	bookmarksnumbered,
	citecolor={nicered},
	linkcolor={myDarkBlue},
	urlcolor={tealgreen},
	pdfstartview=FitH}

\newcommand{\beq}{\begin{equation}}
\newcommand{\eeq}{\end{equation}}
\newcommand{\bseq}{\begin{subequations}}
	\newcommand{\eseq}{\end{subequations}}

\usepackage{times}

\graphicspath{{Figs/}}

\usepackage{orcidlink}
\def\idbako{\orcidlink{0000-0002-3012-6144}}
\def\idchat{\orcidlink{0000-0003-4479-2970}}
\def\idnakas{\orcidlink{0000-0002-3522-5803}}

\begin{document}

\title{Compact objects with primary hair in shift and parity symmetric beyond Horndeski gravities}

\author{Athanasios Bakopoulos\idbako}
\email{atbakopoulos@gmail.com}
\affiliation{Physics Department, School of Applied mathematical and Physical Sciences,
	National Technical University of Athens, 15780 Zografou Campus,
	Athens, Greece.}

\author{Nikos Chatzifotis\idchat }
\email{chatzifotisn@gmail.com}
\affiliation{Physics Department, School of Applied mathematical and Physical Sciences,
	National Technical University of Athens, 15780 Zografou Campus,
	Athens, Greece.}

\author{Theodoros Nakas\idnakas }
\email{theodoros.nakas@gmail.com }
\affiliation{Physics Department, School of Applied mathematical and Physical Sciences,
	National Technical University of Athens, 15780 Zografou Campus,
	Athens, Greece.}
 \begin{abstract}
    \noindent In this work, we delve into the model of the shift symmetric and parity-preserving Beyond Horndeski theory in all its generality. We present an explicit algorithm to extract static and spherically symmetric black holes with primary scalar charge adhering to the conservation of the Noether current emanating from the shift symmetry. We show that when the functionals $G_2$ and $G_4$ of the theory are linearly dependent, analytic homogeneous black-hole solutions exist, which can become regular by virtue of the scalar charge contribution. Such geometries can easily enjoy the preservation of the Weak Energy Conditions, elevating them into healthier compact objects than most hairy black holes in modified theories of gravity. Finally, we revisit the concept of disformal transformations as a solution-generating mechanism and discuss the case of generic $G_2$ and $G_4$ functionals. 
\end{abstract}

\maketitle

\flushbottom

\tableofcontents

\section{Prologue}

The ``no-hair" theorem, also known as the ``uniqueness theorem" in the context of black holes (BH) \cite{Bekenstein:1971, Bekenstein:1972ny, Teitelboim:1972, PhysRevD.51.R6608, Mayo:1996mv, Mazur:2000pn, Chrusciel:2012jk,Hod:2011aa,Ghosh:2023kge}, is a fundamental concept in General theory of Relativity (GR). This theorem essentially states that all black holes can be described by just three externally observable quantities, regardless of their initial conditions or the nature of the matter that formed them. These quantities are: the total mass $M$ of the black hole, the angular momentum $J$, representing the rotation or spin of the black hole,
and the electromagnetic charge $(Q,P)$ of the black hole, although astrophysical black holes are generally considered to be nearly electromagnetically neutral.

The ``no-hair" theorem suggests that these three parameters uniquely define the entire spacetime geometry of a black hole. In other words, any black hole with the same values of mass, angular momentum, and electric charge will have the same external properties, such as its event horizon and gravitational effects, regardless of its formation process or initial conditions. 
This simplicity is metaphorically referred to as a black hole having ``no hair", indicating its characterization by only a limited set of observable quantities.

General Relativity, however, is commonly acknowledged as an effective theory applicable only within the realm of low energies. 
A plethora of cosmological observations indicates instances where GR exhibits limitations, with the most notable challenges being the Dark Energy Problem and GR's inability to account for the inflationary epoch in our universe.
Consequently, such observations motivate us to explore modified gravitational theories.
Over the last few decades, numerous such theories have been formulated; however, among the most elementary and extensively studied ones are the scalar-tensor (ST) theories.
These theories introduce an extra scalar degree of freedom through the presence of a scalar field, which can be either minimally or non-minimally coupled with gravity.
Given that a considerable number of modified and higher-dimensional gravitational theories reduce to scalar-tensor theories under specific conditions, ST theories offer a highly conducive framework for exploring innovative concepts and novel spacetime configurations.
Among ST theories, the most comprehensive theory featuring a single real scalar field and producing second-order field equations is Horndeski theory \cite{Horndeski:1974wa}. 
Recent extensions of Horndeski theory have been devised to handle higher-order field equations without introducing undesirable ghost degrees of freedom.
A particular example of such a theory is Beyond Horndeski theory \cite{Gleyzes:2014dya,Kobayashi:2019hrl}, which will also occupy us in this article.

In a manner analogous to GR, no-scalar-hair theorems \cite{Bekenstein:1972ny, Teitelboim:1972, PhysRevD.51.R6608} have also been developed for ST theories.
These theorems, under certain conditions, impose constraints that prevent black holes from possessing scalar hair---an additional physical quantity arising from the existence of the scalar field in the theory.
However, shortly after their formulation, it became evident that these no-scalar-hair theorems could be evaded. 
Consequently, an abundance of black holes with scalar hair emerged in the scientific literature. 
These include black holes in ST gravities \cite{Bechmann:1995sa, Nucamendi:1995ex, Dennhardt:1996cz, Martinez:2004nb, Bronnikov:2005gm, Nikonov_2008, Anabalon:2012ih, Babichev:2013cya, Anabalon:2013oea, Barrientos:2022avi,
Barrientos:2023tqb,
Sotiriou:2013qea, Sotiriou:2014pfa, Cadoni:2015gfa, Kleihaus:2015iea, Sotiriou:2015pka, Charmousis:2015aya, Benkel:2016rlz, BenAchour:2018dap, Babichev:2020qpr, Karakasis:2021rpn, Bakopoulos:2021dry, Bakopoulos:2022csr, Bakopoulos:2023hkh}, in Gauss-Bonnet and Chern-Simons gravities \cite{Kanti:1995vq, Antoniou:2017acq, Doneva:2017bvd, Silva:2017uqg, Antoniou:2017hxj, Kanti:2019upz, Bakopoulos:2019fbx, Hod:2019pmb, Brihaye:2018grv, Bakopoulos:2018nui, Yagi:2012ya, Cano:2019ore, Chatzifotis:2022mob, Chatzifotis:2022ene,  Filippini:2019cqk}, black holes under Abelian and non-Abelian vector fields \cite{Torii:1993vm, Gubser:2005ih, Anabalon:2013qua, Sanchis-Gual:2016tcm, Hod:2017kpt, Pacilio:2018gom, Zou:2019ays, Fernandes:2020gay, Brihaye:2021ich, Hong:2020miv,  Astefanesei:2019mds,  Astefanesei:2020xvn,  Myung:2020ctt, Martinez:2006an,  Babichev:2015rva,  Fan:2015oca, Guo:2021zed, Heisenberg:2017xda, Brihaye:2019gla, Santos:2020pmh,Karakasis:2022xzm, Mavromatos:1995kc, Mavromatos:1997zb, Mavromatos:1999hp}, and other local solutions stemming from theories of higher or lower spacetime dimensionality \cite{Babichev:2023rhn, Fernandes:2021dsb, Babichev:2023dhs, Brihaye:2018woc, Ovalle:2020kpd, Faraoni:2021nhi, Karakasis:2021lnq, Karakasis:2021ttn, Karakasis:2022fep}.

It is important to clarify at this point, that the scalar hair characterizing a black hole can be distinguished in \emph{primary} and \emph{secondary}.
In the context of a ST theory, primary scalar hair/charge can be defined as the physical quantity that characterizes the black hole in addition to its mass, angular momentum, and electric charge.
In this case, the scalar hair/charge is directly related to an internal symmetry of the theory which consequently leads to the existence of a Noether current \cite{Bakopoulos:2023fmv}.
Especially within the framework of beyond Horndeski theory, in Sec. \ref{sec:gen-frame}, we will see precisely how the primary charge is defined and evaluated.
On the other hand, the secondary hair refers to modifications in the black hole's metric, where despite the inclusion of additional non-trivial fields beyond electromagnetism, the entire configuration is still entirely determined by the black hole's mass, angular momentum, and electric charge.

Until recently, no explicit black holes with primary hair had been discovered in single-field scalar-tensor theories. 
A few other examples exist in the context of other theories \cite{Herdeiro:2014goa, Charmousis:2014zaa, Herdeiro:2015gia}, while several black holes with secondary scalar hair have been more easily constructed, bypassing the no-scalar-hair theorem(s). In general, the complexity of scalar-tensor theories seldomly allows for exact analytical local solutions. However, by imposing additional symmetries in the underlying
gravitational action, exact solutions may be derived. This is the case for shift-symmetric and parity-preserving beyond Horndeski theory. In such a framework, the fact that the action depends solely on derivatives of the scalar field allows the scalar field to be linearly dependent on the time coordinate while preserving compatibility with a static metric ansatz. Making use of this concept, the existence of black holes with primary scalar hair in shift-symmetric and parity-preserving beyond Horndeski gravities was recently proven \cite{Bakopoulos:2023fmv}. Following on this, black holes with primary hair were also studied in \cite{Baake:2023zsq} for arbitrary spacetime dimensions, where Maxwell fields and Lovelock corrections in the action were also considered.

In this work, we are also focusing on the framework of the shift-symmetric and parity-preserving beyond Horndeski theory in all its generality. We present a way to extract homogeneous static and spherically symmetric black-hole solutions with primary charge in a semi-agnostic subclass of beyond Horndeski gravity, described by a linear dependence of the $G_2$ and $G_4$ functionals under the reasonable assumption that the theories yield a smooth limit to General Relativity. 
We prove that such solutions can always become regular at a critical value of the black hole ADM mass by virtue of the primary charge contribution. Moreover, we have verified that such configurations can always be fixed to respect the Weak Energy Conditions, thus elevating our local solutions to healthier compact objects in comparison to most hairy black holes in Horndeski gravity. An important result of this analysis is that in the generic subclass that we are working on, a canonical kinetic term in the action will in general always yield solid deficit angles. We have shown that in order to avoid such pathologies, a pure disformal transformation is sufficient to result in inhomogeneous configurations with correct Minkowski asymptotics. Motivated by this result, we discuss the case of wormhole solutions in the generic beyond Horndeski gravity sourced by the primary charge and present indicative arguments that such solutions cannot naturally exist if the seed action is well-defined, i.e. possesses a correct GR limit. Finally, we delve into the case of generic $G_2$ and $G_4$ functionals. Due to the sheer number of degrees of freedom of the problem, we focus on theories yielding homogeneous solutions and fix the kinetic term of the scalar field, $X=-\frac{1}{2}\partial_\mu\Phi\partial^\mu\Phi$, to be expressed as a polynomial equation of the radial coordinate, having also a smooth limit to GR vacuum solutions. 
To this end, we are primarily focused on the strength of the algorithm presented here, which can be used to obtain new explicit local solutions, rather than examining in detail the solutions.

This work is organized as follows: In Sec. \ref{sec:gen-frame}, we outline the general theoretical framework on which this article is focused. 
In Sec. \ref{sec:hom-sols} and \ref{disformal-section}, we investigate homogeneous black-hole solutions and disformal transformations respectively. 
Sec. \ref{nolinear} extends the methodology to more generic beyond Horndeski theories, while the conclusions and future directions of research are presented in Section \ref{conclusion}.

\section{The general framework}\label{sec:gen-frame}

We begin our analysis by firstly providing the generic shift-symmetric and parity-preserving beyond Horndeski theory under consideration, whose action in geometrized units ($c=G=1$) reads
\begin{equation}
    \label{eq:act}
    S=\int \frac{d^4x\sqrt{|g|}}{16\pi}\left[G_4(X)R+G_{4X}[(\square\Phi)^2-\Phi_{;\mu\nu}\Phi^{;\mu\nu}]+G_2(X)+F_4(X)\epsilon^{\mu\nu\rho\sigma}\epsilon^{\alpha\beta\gamma}_{\,\,\,\,\,\,\,\,\,\sigma}\Phi_{;\mu}\Phi_{;\alpha}\Phi_{;\nu\beta}\Phi_{;\rho\gamma}\right]\,,
\end{equation}
where $\Phi_{;\mu}\equiv\partial_\mu\Phi$, $\Phi_{;\mu\nu}\equiv\nabla_\mu \partial_\nu \Phi$, $\displaystyle X\equiv-\frac{1}{2}\partial^\mu\Phi\partial_\mu\Phi$ is the kinetic term of the scalar field, and
\begin{equation}
    \label{eq:phi}
    \Phi(t,r)=qt+\Psi(r)\,.
\end{equation}
Note that the linear time dependence in the expression of the scalar field $\Phi$ is allowed due to the shift symmetry of the considered Lagrangian density, while the parameter $q$ has dimensions $[L]^{-1}$, where $[L]\equiv$ (length units), since the scalar field is dimensionless.
Also, the internal shift symmetry of the theory \eqref{eq:act} results in the existence of a Noether current which is given by
\begin{equation}
    \label{eq:Noether}
    \bold{J}=J_\mu \diff x^\mu,\hspace{1em}J^{\mu}=\frac{1}{\sqrt{|g|}}\frac{\delta S}{\delta (\del_\mu \Phi)}\,,
\end{equation}
while the equation of motion regarding the scalar field is expressed as $\nabla_\mu J^\mu=0$.
This results in a primary scalar charge which is related to the parameter $q$ and is given by
\begin{equation}
    \label{eq:sc-char-def}
    Q_s=\frac{1}{N}\int \star \bold{J}\,,
\end{equation}
where the $\star$ operator is the Hodge dual, $N$ is a normalization constant and $J^r=0$.

Throughout the current article, we will be occupied by static and spherically symmetric spacetime configurations, hence, the line element is of the form
\begin{equation}
    \label{eq:ds}
    ds^2=-h(r) \diff t^2+\frac{1}{f(r)}\, \diff r^2+r^2 \diff \Omega^2\,,
\end{equation}
with $\diff \Omega^2=\diff\theta^2+\sin^2\theta\, \diff\phi^2$.
The variation of the action functional \eqref{eq:act} with respect to the metric tensor and the scalar field leads us to the field equations of the theory.
Utilizing the results of \cite{Babichev:2015rva, Bakopoulos:2022csr,Bakopoulos:2023fmv}, for the metric ansatz \eqref{eq:ds} and the expression \eqref{eq:phi} for the scalar field, the resulting independent field equations can be brought to the following form:
\begin{gather}
    \frac{f}{h}=\frac{\gamma^2}{Z^2}, \label{eq1}\\
    r^2(G_2 Z)_{X}+2(G_4 Z)_X \left(1-\frac{q^2 \gamma^2}{2 Z^2 X}\right)=0,\label{eq2}\\
    2\gamma^2 \left(h r-\frac{q^2 r}{2 X}\right)'=-r^2 G_2 Z-2G_4 Z\left(1-\frac{q^2 \gamma^2}{2 Z^2 X}\right)+\frac{q^2\gamma^2X' r}{ZX^2}\left( 2XG_{4X}-G_4\right)\label{eq3}\,.
\end{gather}
In the above, prime denotes differentiation  with respect to $r$, the subscript $X$ denotes differentiation  with respect to the kinetic term, while $Z\left(X\right)$ is an auxiliary function which allows us to write the field equations in a more compact way and is defined as
\begin{equation}
    \label{eq:Z}
    Z\left(X\right)\equiv 2XG_{4X}-G_4(X)+4X^2F_4(X)\,.
\end{equation}
For the derivation of the above equations we also used the relation 
\begin{equation}\label{xrep}
X=\frac{1}{2}\left(\frac{q^2}{h}-f\Psi'^2\right).
\end{equation}
Therefore, given the theory and the functional form of $G_2(X)$, $G_4(X)$, and $F_4(X)$, the above equations can be solved for the three unknown functions $h(r)$, $f(r)$, and $\Psi(r)$.

Focusing on eq. \eqref{eq2}, we observe that it is an algebraic equation for $X$, while (\ref{eq3}) is a first-order differential equation for the metric function $h(r)$. This implies that for most choices of $G_2, G_4$, and $F_4$, the system can yield a solution, even in an integral form. However, selecting arbitrary forms for the aforementioned functions does not guarantee that the solution will possess the desired characteristics to describe a compact object. Nevertheless, the form of the equations enables us to derive a straightforward algorithm regarding the selection of coupling functions and the construction of solutions.
Before we proceed with the description of the algorithm, we note that the choice 
\begin{equation}\label{jgzcond}
   (G_2 Z)_X=\Xi\, (G_4 Z)_X ,
\end{equation}
with $\Xi=\Xi(X)$ a new auxiliary function, brings eq. \eqref{eq2}  in a much more meaningful form, namely
\begin{equation}\label{eq2rev}
    X Z^2(2+r^2 \Xi)-q^2\gamma^2=0\,.
\end{equation}
The algorithm consists of the following steps. Firstly, the decision is made on whether the solution will be homogeneous or not. From eq. (\ref{eq1}), it is observed that a constant value of $Z$ leads to a homogeneous solution, while a nontrivial form of $Z$ results in a non-homogeneous solution. Subsequently, the function $\Xi$ is chosen to solve eq. (\ref{eq2rev}) with respect to $X$. The selection of $\Xi$ ensures that $X$ exhibits the desired characteristics.
Once $Z$ and $\Xi$ are chosen, a form for $G_2$ is selected, and eq. (\ref{jgzcond}) yields $G_4$ (or vice versa). Finally, eq. (\ref{eq3}) is solved to find the functional form of $h$. 

Throughout this work, we will mainly focus on theories that lead to the following expression for the kinetic term of the scalar field
\begin{equation}
    X=\frac{\beta q^2}{2\beta+\alpha r^2},
\end{equation}
while the generic case will be tackled in section \Ref{nolinear}.
The constants $\beta$ and $\alpha$ are related to the coupling constants of the theory. 
The specific form of $X$ exhibits constant asymptotics at both $r=0$ and infinity, a characteristic commonly associated with well-behaved solutions for compact objects. Moreover, this particular selection typically results in analytical solutions for the metric function $h(r)$. Additionally, it is noteworthy that constructing a non-homogeneous solution is always possible by applying a disformal transformation to a homogeneous solution (see Section \ref{disformal-section}). Consequently, our emphasis will predominantly be on the scenario where $Z$ is a constant, leading to homogeneous solutions.

\section{Homogeneous solutions}\label{sec:hom-sols}

For simplicity, we start our analysis by setting $\Xi$ to be constant. Pertaining now to homogeneous solutions, i.e. $h(r)=f(r)$ and $Z=\gamma$, it is interesting to observe that the proportionality between $G_4(X)$ and $G_2(X)$ makes eq. \eqref{eq2rev} trivially solvable for $X$ (this property is also discussed in \cite{Baake:2023zsq} for arbitrary dimensions).\,\footnote{For non-homogeneous solutions with primary scalar hair in the context of shift-symmetric and parity-preserving pure Horndeski theory with $\Xi=const.$ the reader is referred to Appendix \ref{ap:horn-sols}.}
Hence, by assuming $G_2=\alpha S(X)$ and $G_4(X)=\zeta+\beta S(X)$, eq. \eqref{eq2rev} \textit{always} yields to $X=\frac{\beta q^2}{2\beta+\alpha r^2}$. 
In order to be sure that $X$ is everywhere regular, we set $\alpha=2\beta/\lambda^2$ obtaining
\begin{equation}
    \label{eq:X-hom-prop}
    X=\frac{q^2}{2}\frac{1}{1+(r/\lambda)^2}\,.
\end{equation}
Given the action \eqref{eq:act}, it is straightforward to deduce that the function $G_2$ has dimensions $[L]^{-2}$, $[G_4]=[Z]=1$, while $[F_4]=[L]^4$.
Consequently, by assuming that the function $S(X)$ is dimensionless, we can readily infer the dimensionality of the coupling constants, i.e. $[\alpha]=[L]^{-2}$, $[\zeta]=[\beta]=[\gamma]=1$, and $[\lambda]=[L]$\,.
We can now directly integrate \eqref{eq3} and obtain the expression for the metric function $h(r)$, which is given by
\begin{equation}
    \label{eq:h-hom-prop}
    h(r)=1+\frac{C}{r}+\left(1+\frac{\zeta}{\gamma}\right)\frac{r^2}{\lambda^2}+\frac{2\beta}{\gamma \lambda^2}\frac{1}{r}\int r^2 (S-2XS_X) \diff r\,.
\end{equation}
In order to have a well-defined theory, it is natural to assume that the function $S(X)$
is analytic in $X$. To further generalize the framework, we extend upon the notion of analytic functions and incorporate the cases that are finite at $X\rightarrow0$ (when $q\rightarrow 0$), but not necessarily differentiable at that point. To this end, we can express the function $S(X)$ in a power series in the following way:
\begin{equation}
    \label{eq:S-exp}
    S(X)=\sum_{n=0}^\infty c_{\frac{n}{s}} X^{\frac{n}{s}}\,,\hspace{1em} s\in \mathbb{Z}^+\,,
\end{equation}
where in accordance to the preceding discussion, $[c_{\frac{n}{s}}]=[L]^{\frac{2n}{s}}$.
We have included the parameter $s$ to allow for the examination of any positive rational exponent of $X$, something that will prove useful below.
Note also that one may consider finite terms in the above expansion by selectively fixing the desired constants $c_{n/s}$ to zero.
By substituting \eqref{eq:S-exp} into \eqref{eq:h-hom-prop} and utilizing the interchangeability of summation and integration, we are led to 

\begin{align}
    h(r)={}&{}1+\frac{C}{r}+\left(1+\frac{\zeta}{\gamma}+\frac{2\beta}{3\gamma}\, c_0\right)\frac{r^2}{\lambda^2}+\frac{2\beta}{3\gamma} \frac{r^2}{\lambda^2} \sum_{n=1}^{\infty} c_{\frac{n}{s}} \left(1-\frac{2n}{s} \right) \left(\frac{q^2}{2} \right)^{n/s} \,_2F_1\left(\frac{3}{2},\frac{n}{s};\frac{5}{2};-\frac{r^2}{\lambda^2} \right)\,.
    \label{eq:h-hom-prop-F}
\end{align}
In the above, the expression of the hypergeometric function with a negative argument and an absolute value greater than unity may initially appear erroneous. 
However, it actually constitutes a more concise representation of a rigorously defined hypergeometric function, as it adheres to the following relation
$$
\,_2F_1\left(\frac{3}{2},\frac{n}{s};\frac{5}{2};-\frac{r^2}{\lambda^2} \right)=\bigg(1+\frac{r^2}{\lambda^2}\bigg)^{-n/s}\,_2F_1\left(\frac{n}{s},1;\frac{5}{2};\frac{1}{1+\lambda^2/r^2} \right)\,.$$
Note also, that the preceding relation stems from the Pfaff transformation
$$
\,_2F_1\left(a,b;c;z\right)=(1-z)^{-b}\,_2F_1\left(b,c-a;c;\frac{z}{z-1}\right)\,.$$

In the asymptotic regime, $r\rightarrow+\infty$, the metric function takes the form
\begin{align}
    h(r)=1&+\frac{1}{r}\left[C+\frac{\lambda\beta\sqrt{\pi}}{2\gamma} \sum_{n=1}^\infty c_{\frac{n}{s}} \left(1-\frac{2n}{s} \right) \left(\frac{q^2}{2} \right)^{n/s} \frac{\Gamma\left(\frac{n}{s}-\frac{3}{2}\right)}{\Gamma\left(\frac{n}{s}\right)}\right]+\left(1+\frac{\zeta}{\gamma}+\frac{2\beta}{3\gamma}\, c_0\right)\frac{r^2}{\lambda^2}\nonumber\\[1mm]
    &+\frac{2\beta}{3\gamma}  \sum_{n=1}^\infty c_{\frac{n}{s}} \left(\frac{q^2}{2} \right)^{n/s}\left(\frac{\lambda}{r}\right)^{2n/s}\left[\left(\frac{1-\frac{2n}{s}}{1-\frac{2n}{3s}}\right)\frac{r^2}{\lambda^2}-\frac{3n}{s}+\mathcal{O}\left(\frac{1}{r^2}\right)\right]\,.
    \label{eq:h-hom-prop-exp-infty}
\end{align}
It is obvious from the above expression that the second term will contribute to the ADM mass of the compact object, while an additional contribution to the ADM mass comes from the bottom-line summation in the case where $2n/s=1$ .
The third term decides whether the spacetime is asymptotically flat or not.
In the context of this article, we are interested in asymptotically flat compact-object solutions and therefore henceforth we set $c_0=0$ and $\zeta=-\gamma$ to always have Minkowski asymptotics. 
In addition to what we already mentioned, one has to verify that the bottom-line summation does not generate global-monopole terms \cite{Barriola:1989hx,Chatzifotis:2022ubq} and that the ADM mass is finite everywhere. Consequently, one has to set $c_1=0$ in order to make the monopole configurations to vanish from the last summation.
As for the terms in the first summation, one has to cancel all the terms---through the constants $c_{n/s}$---that make the Gamma functions blow up.
Moreover, from the expansion of the metric function at $r\rightarrow 0$, i.e.
\begin{equation}
    h(r)=1+\frac{C}{r}+\frac{2\beta}{3\gamma} \frac{r^2}{\lambda^2} \sum_{n=1}^{\infty} c_{\frac{n}{s}} \left(1-\frac{2n}{s} \right) \left(\frac{q^2}{2} \right)^{n/s} \left[1-\frac{3n}{5s}\frac{r^2}{\lambda^2}+\mathcal{O}(r^4) \right]\,,
    \label{eq:h-hom-prop-exp-0}
\end{equation}
it is straightforward to deduce that in order to obtain regular black-hole and soliton configurations, it is mandatory to set $C=0$, which, as previously mentioned, does not cancel out the effective ADM mass term. 

Let us now evaluate the scalar charge that accompanies the solutions described by the metric function \eqref{eq:h-hom-prop-F}. 
To this end, we need first to compute the components of the Noether current $J^\mu$ and then make use of the defining relation \eqref{eq:sc-char-def} for the scalar charge.
By doing so, we obtain
\begin{gather}
    \label{eq:J-hom-prop}
    J^\mu=\left(-\frac{4\beta q}{r^2+\lambda^2}\, S_X,0,0,0\right)\,,\\[2mm]
    \label{eq:Qs-hom-prop}
    Q_s=\frac{1}{N}\int \star \bold{J}=\frac{4\pi}{N} \int  r^2 J^t\, \diff r=\frac{8\pi^{3/2}\beta}{N}\frac{\lambda}{q}\sum_{n=1}^\infty c_{\frac{n}{s}}\frac{n}{s}\left(\frac{q^2}{2}\right)^{n/s}\frac{\Gamma\left(\frac{n}{s}-\frac{3}{2}\right)}{\Gamma\left(\frac{n}{s}\right)}\,, \hspace{1em} \forall\,\frac{n}{s}>\frac{3}{2}.
\end{gather}
It is crucial to observe that both the asymptotic expansion of the metric function and the expression for the scalar charge yield identical constraints on the permissible values for the parameter $n/s$. Examining \eqref{eq:h-hom-prop-exp-infty} and \eqref{eq:Qs-hom-prop}, it is evident that both the ADM mass and the scalar charge $Q_s$ become undefined for those instances of $n/s$ where the Gamma functions reach infinity. Consequently, regardless of the chosen value for $s$, any constant $c_{n/s}$ with $n/s \leq 3/2$ must be set to zero.

Having established the permissible non-zero values for the constants $c_{n/s}$ within a given ansatz, we can proceed to examine specific examples that hold particular significance. 
To begin, when $s=1$, the function $S(X)$, as provided in \eqref{eq:S-exp}, represents the expansion of an analytic function in terms of $X$, thus, for any given function,
the solution will invariably be described by \eqref{eq:h-hom-prop-F}.
Also, according to the preceding discussion, we know that in every case, the term $c_1 X$ must be excluded from the expansion.
For example, one can readily consider a theory that is characterized by the function $S(X)=aX-e^{a X}$, or any other similar combination of a canonical kinetic term and an analytic function.
Furthermore, considering cases with $s\neq 1$, the theory can be naturally extended.
Notably, the case $s=2$ holds particular interest because, in addition to its capacity to describe any analytic function, it incorporates terms of the form $X^{\frac{2k+1}{2}}$ for $k\in \mathbb{Z}^{\geq}$.
What is interesting here is the fact that for all these terms, the hypergeometric function in \eqref{eq:h-hom-prop-F} will take the form $\,_2F_1\left(\frac{3}{2},\frac{2k+1}{2};\frac{5}{2};-\frac{r^2}{\lambda^2} \right)$ which can always be written in a closed form.

Before we investigate the case $s=2$ more thoroughly, it is important to note that compact-object solutions described by \eqref{eq:h-hom-prop-F} can easily respect the Weak Energy Conditions (WEC), as opposed to many hairy black-hole solutions in modified gravities. 
Indeed, considering a normalized timelike vector $l$, the WEC read
\begin{equation}
    T_{\mu\nu}l^\mu l^\nu=-\frac{2\beta c_0+3 (\gamma +\zeta )}{\gamma  \lambda ^2}+
    \frac{2\beta}{\gamma\lambda^2}\sum_{n=1}^{\infty}c_{\frac{n}{s}}\left(\frac{q^2}{2}\right)^{\frac{n}{2}}\left(\frac{2n}{s}-1\right)\left(1+\frac{r^2}{\lambda^2}\right)^{-\frac{n}{s}}
\end{equation}
As can be immediately verified, the Weak Energy Conditions will be respected for such configurations, assuming an appropriate fixing of the coupling constants in the action. Indeed, omitting the cosmological constant $c_0$ and setting $\zeta=-\gamma=1$ to yield correct Minkowski asymptotics and a unitary coupling to the Ricci scalar, the positivity of the above expression can be easily enforced by the choice $\beta\sum_{n=1}^{\infty}c_{\frac{n}{s}}<0$ under $s\leq2$. The reasoning behind that is that such configurations are dominated by the tangential pressure of the effective perfect fluid. This lies in agreement with the findings of \cite{Dorlis:2023qug}, where it was shown that there are two distinct ways to bypass the violation of the energy conditions for hairy and spherically symmetric black holes, depending on the homogeneity of the geometry. Our result falls in the first case, where the radial pressure of the perfect fluid is dominated by the corresponding tangential pressure, while the second case applies only to inhomogeneous black holes.

\begin{figure}[t]
    \centering
    \begin{subfigure}[b]{0.49\textwidth}
    \includegraphics[width=1\textwidth]{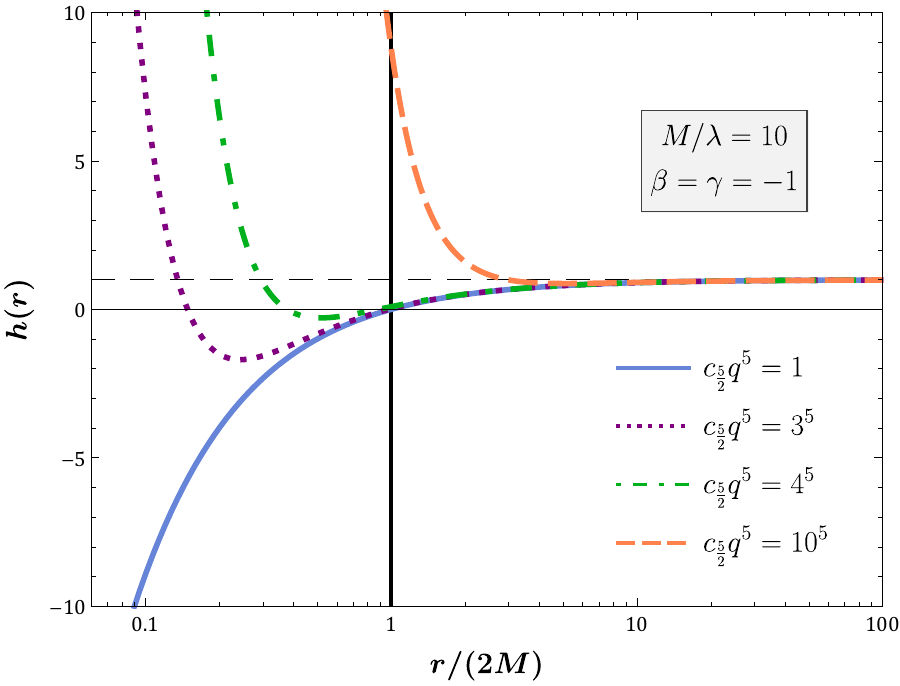}
    \caption{\hspace*{-1.5em}}
    \label{subf:sbh}
    \end{subfigure}
    \hfill
    \begin{subfigure}[b]{0.495\textwidth}
    \includegraphics[width=1\textwidth]{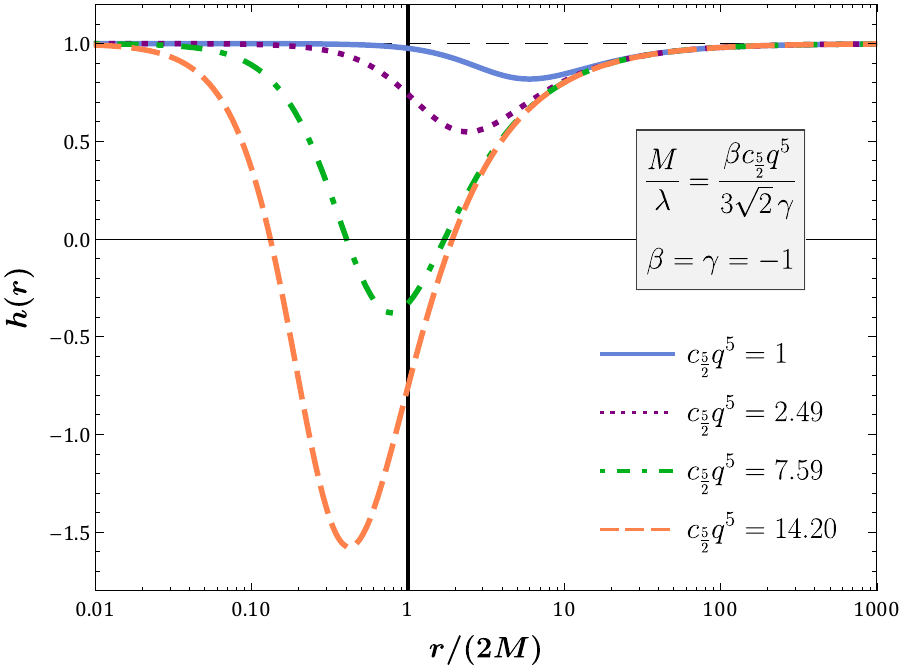}
    \caption{\hspace*{-1.7em}}
    \label{subf:rbh}
    \end{subfigure}
    \vspace*{-0.5em}
    \caption{(a) One single-horizon black hole, two black holes with two horizons, and a naked singularity. (b) Two regular black-hole and two soliton solutions.
    The horizontal axis in both figures is logarithmic.}
    \label{fig:sols}
\end{figure}

Let us now delve into the $s=2$ case. From the preceding discussion we already know that in order to have an asymptotically flat solution with well-defined ADM mass and scalar charge, we need to impose $\zeta=-\gamma$ and $c_0=c_{1/2}=c_1=c_{3/2}=0$.
The first non-zero term in the expansion \eqref{eq:S-exp} corresponds to $n/s=2$, for which one finds that
\begin{equation}
    h(r)=1+\frac{C}{r}-\frac{3\beta c_2 q^4}{4\gamma}\left[\frac{\arctan(r/\lambda)}{r/\lambda}-\frac{1}{1+(r/\lambda)^2} \right]\,.
\end{equation}
This black-hole solution was recently discovered in \cite{Bakopoulos:2023fmv}, representing the inaugural instance of a solution featuring primary scalar hair in beyond Horndeski theory.
Hence, to obtain a new solution we have to go beyond $n/s=2$.
The next term that we can consider is $n/s=5/2$ for which we obtain
\begin{equation}
    \label{eq:h-5/2-m}
    h(r)=1-\frac{2M}{r}+\frac{\sqrt{2}\,\beta c_{\frac{5}{2}}  q^5}{3 \gamma }\frac{\lambda}{r}-\frac{\sqrt{2}\, \beta c_{\frac{5}{2}} q^5}{3 \gamma}\frac{r^2/\lambda^2 }{\left(1+r^2/\lambda^2\right)^{3/2}}\,.
\end{equation}
Note that in the aforementioned expression, we have chosen $C=-2M+\sqrt{2}\beta c_{\frac{5}{2}} \lambda q^5/(3 \gamma)$ to ensure that the parameter $M$ corresponds to the ADM mass. 
Notice that by choosing $M/\lambda=\beta c_{\frac{5}{2}}  q^5/(3\sqrt{2} \gamma)$ the $1/r$ term in the above expression vanish identically.
In this scenario, the solutions described by the metric function $h(r)$ could vary from regular black holes to solitons, as shown in Fig.\,\ref{subf:rbh}.
It is also important to mention that in this case, our solution mimics at a very good extent the Bardeen solution \cite{Bardeen} of non-linear electrodynamics. 
On the other hand, allowing the $1/r$ term results in solutions ranging from singular black holes to naked singularities, see Fig.\,\ref{subf:sbh}.
All parameters depicted in Fig.\,\ref{fig:sols} are dimensionless.
Finally, notice that for the chosen values of the parameters, the horizon radii of the singular black holes (Fig.\,\ref{subf:sbh}) reside at $r_h\approx 2M$, while the regular black holes (Fig.\,\ref{subf:rbh}) have $r_h> 2M$. 
This means that the regular black-hole solutions have a larger horizon radius than the corresponding Schwarzschild black hole of the same mass.
Given the values of the theory parameters, one can readily deduce whether the resulting black hole has a greater or lower horizon radius compared to that of a Schwarzschild black hole, by simply substituting the value $r=2M$ in the expression \eqref{eq:h-5/2-m} and determine the sign of $h(2M)$.
By doing so, one finds that
\begin{equation}
    \label{eq:h2M}
    h(2M)=\frac{\beta c_{\frac{5}{2}} q^5}{3\sqrt{2}\,\gamma} \frac{\lambda}{M} \left[1-\frac{8M^3}{\lambda^3}\left(1+\frac{4M^2}{\lambda^2}\right)^{-3/2} \right]\,.
\end{equation}
A black-hole solution with $h(2M)>0$ will be more sparse than the corresponding Schwarzschild black hole of the same mass, while for $h(2M)<0$ the resulting solution will be more compact than the Schwarzschild one. We note that 
 one may easily verify the existence of a critical value $m_c\approx 0.393076$ such that $h(2M)=0$ for $M/\lambda=m_c $, $h(2M)>0$ for $M/\lambda<m_c $ and $h(2M)<0$ for $M/\lambda>m_c $.

A key insight drawn from the preceding analysis is that within the realm of shift-symmetric and parity-preserving beyond Horndeski theories, particularly in cases where the functions $G_4(X)$ and $G_2(X)$ exhibit proportionality, we have successfully devised a comprehensive algorithm that enables the explicit derivation of compact-object solutions with primary scalar hair/charge for any theory governed by the function $S(X)$ in eq.\,\eqref{eq:S-exp}.

\section{Disformal Transformations}\label{disformal-section}

A disformal transformation of the spacetime metric tensor is a well-established process that transforms an initial Horndeski solution into a solution that belongs to the beyond Horndeski framework. Additionally, a disformal transformation applied to a theory already classified as beyond Horndeski results in its mapping to an alternative beyond Horndeski class. In the subsequent analysis, we consider a specific scenario, where quantities denoted with a hat symbol represent the disformed functions, while those without the hat symbol signify the initial ``seed" solution. In this context, we denote the seed variables as $\Phi$, $h$, $f$, and, notably, $X$, within the context of a specific (beyond) Horndeski theory characterized by ${G_2, G_4, F_4}$. The disformal transformation is explicitly expressed as follows:
\begin{equation}
    \hat{g}_{\mu\nu}=g_{\mu\nu}-W(X)\partial_\mu\Phi\partial_\nu\Phi. 
\end{equation}
Note that under a disformal transformation, the scalar field remains invariant $\hat \Phi= \Phi$.  
Also, since the disformal mapping does not contain any explicit $\Phi$ dependence,
the shift symmetry is retained in the target theory.
By applying the transformation to the seed metric (\ref{eq:ds}), we get 
\begin{equation}
 d\hat{s}^2=-\left(h+q^2 W \right)dt^2-2q \Psi' W dt dr+\left(\frac{1-Wf\Psi'^2}{f}\right)dr^2+r^2d\Omega^2. 
\end{equation}
Redefining the time coordinate as $dt=d\tau-\frac{q \Psi' W}{h+q^2 W} dr$,
 we can eliminate the cross terms that appear in the above metric.  Using this new coordinate, the metric takes the following form
 \begin{equation}\label{dismet}
    d\hat{s}^2=-(h+q^2W)d\tau^2+\frac{Z^2}{\gamma^2}\frac{1+2 WX}{(h+q^2W)}dr^2+r^2d\Omega^2
\end{equation}
where in the derivation of the above equation we have used eqs. (\ref{eq1}) and (\ref{xrep}). Note that although the scalar field does not change under the disformal transformation, it will change due to the coordinate transformation. Therefore, in the $\tau$ coordinate the scalar field is 
\begin{equation}
    \Phi=q\tau + \Psi - q^2\int\frac{ \Psi' W}{h+q^2 W} dr.
\end{equation}
The field is still linear in the new time coordinate since the shift symmetry remains. 

From eq. (\ref{dismet}), it is evident that under a disformal transformation, a homogeneous seed solution undergoes a transformation into a non-homogeneous one. This flexibility in transforming solutions is a notable advantage of disformal transformations, providing a versatile tool for exploring diverse spacetime structures. By strategically choosing a disformal function, specifically such that $1+2W X=\gamma^2/Z^2$, we consistently arrive at a homogeneous local solution within the framework of shift and parity symmetric Beyond Horndeski theory. This characteristic not only underscores the power of disformal transformations in generating homogeneous solutions but also establishes their efficacy within the specific context of Beyond Horndeski theories, particularly those adhering to shift and parity symmetry. Moreover, it is noteworthy that the resulting metric, after such transformations, will generally exhibit an explicit dependence on $q$,  while the disformed action does not contain such a dependence, thereby confirming the existence of a primary charge.   

\subsection{Regularization of monopole-like configurations}

Beyond the transformation of homogeneity, disformal transformations offer a systematic approach to manipulating metric structures, providing a valuable means to study gravitational theories comprehensively. The ability to control and modify solutions through disformal transformations enhances our capacity to model various astrophysical phenomena and explore a broader range of theoretical scenarios. Notably, from eq. (\ref{dismet}), it becomes evident that, due to the linear time dependence of the scalar field, the disformed metric contains terms of the form $q^2W$. Consequently, through the application of the appropriate disformal transformation, it becomes feasible to rectify metric pathologies, such as the solid angle deficits observed in the solutions of the previous section. 

As an illustrative example, we will adopt a monopole configuration as our seed solution, which was initially explored in \cite{Bakopoulos:2023fmv}. This particular solution belongs to the class described by eq. (\ref{eq:h-hom-prop-F}), specifically characterized by $c_1\neq 0$, while all other $c_i$ coefficients  trivially vanish. Assuming the same notation with \cite{Bakopoulos:2023fmv}, the
seed configuration is
    \begin{equation}
         h(r)=1+\eta q^2-\frac{2M}{r}+\eta q^2\frac{\pi/2-\arctan(r/\lambda)}{r/\lambda},\qquad f(r)=h(r),\qquad X(r)=\frac{\lambda ^2 q^2}{2 \left(\lambda ^2+r^2\right)}.
    \end{equation}
 The above solution is supported by the seed action of $ G_2=2\frac{\eta}{\lambda^2}X,\, G_4=1+\eta X,$ and $F_4=-\frac{\eta}{4X}$.

Therefore, it is clear that a disformal function of the form  $W(X)=-\eta$, would cancel out the deficit that appears in the seed metric. The transformed solution has the following form
    \begin{equation}
        d\hat{s}^2=-\left(1-\frac{2M}{r}+\eta q^2\frac{\pi/2-\arctan(r/\lambda)}{r/\lambda}\right)d\tau^2+\frac{r^2+\lambda^2-q^2\eta\lambda^2}{(r^2+\lambda^2)\left(1-\frac{2M}{r}+\eta q^2\frac{\pi/2-\arctan(r/\lambda)}{r/\lambda}\right)}dr^2+r^2d\Omega^2.
    \end{equation}

By expanding the solution at infinity, it becomes evident that the new solution is asymptotically flat, with the ADM mass being $M$. 
The existence of the horizon readily confirms that the provided solution characterizes a black hole. Examining the expansion near $r=0$, we deduce that  
\begin{equation}
    |g_{\tau\tau}|=1-\eta  q^2+\frac{\frac{1}{2} \pi  \eta  \lambda  q^2-2 M}{r}+\mathcal{O}(r^2).
\end{equation}
Hence, even when $M=\frac{1}{4} \pi  \eta  \lambda  q^2$, the regularization of the black hole is unattainable since a singularity persists, attributed to the $\eta q^2$ term. 
The numerator of $g_{rr}$ encounters an issue with a problematic root occurring at $r^2=\lambda^2(\eta q^2-1)$. At this radius, the disformal transformation becomes non-invertible. The occurrence of this root can be prevented if $\eta q^2 < 1$ or if $\eta<0$. In the latter case, the seed black hole possesses a phantom kinetic term.
 The detailed investigation of the phenomenology associated with this solution exceeds the scope of the current work, and we defer a comprehensive analysis to a future article.

\subsection{Wormholes in parity-preserving and shift-symmetric theories}

A traversable wormhole, expressed in Schwarzschild-like coordinates, is characterized by the following metric:
\begin{equation}
ds^2=-e^{2A(r)}dt^2 + \frac{dr^2}{1-\frac{B(r)}{r}}+ r^2d\Omega^2,
\end{equation}
where $A(r)$ and $B(r)$ are known as the redshift and shape functions, respectively \cite{Morris:1988cz}. Within this coordinate system, the throat is defined by the equation $B(r_0)=r_0$. It is essential to note that the condition for the absence of horizons necessitates $A$ to be everywhere finite. 
From eq. (\ref{eq1}), it is evident that, under the conditions of parity and shift symmetry, the requirement for the existence of the throat is $\left(1/Z^2\right)\big|_{r=r_0}=0$. The auxiliary function $Z$ is a function of the kinetic term, with the kinetic term itself being determined from the algebraic eq. (\ref{eq2}). It is apparent from this equation that $X(r)$ and, consequently, $Z(r)$ depend solely on the scalar charge $q$ and the coupling constants of the theory. This implies that the throat radius $r_0$, if it exists, will also  depend on these parameters.

In the cases where $q=0$, a No-Go theorem concerning the existence of a wormhole with a mass dependent  throat can be readily derived. This is because the kinetic term of the scalar field $X$ will always be mass independent, as can be immediately verified from (\Ref{eq2}). As such, in this case, the wormhole throat is solely dependent on the coupling constants of the theory. Given that, for a particular theory, these coupling constants are fixed, it follows that, in theories supporting wormholes, the wormhole throat remains constant and does not exhibit variability akin to the horizon throat of black holes like the one presented in \cite{Chatzifotis:2021hpg}. On the other hand, wormholes with a mass dependent throat were found in \cite{Bakopoulos:2021liw} in the context of shift-symmetric, but parity breaking theories. 
 
When $q\neq 0$, the wormhole throat, in addition to depending on the coupling constants, will also be influenced by the scalar charge $q$. As this parameter is an integration constant, the throat radius can theoretically vary. Instead of searching for theories with $Z$ functions that can support both a throat and a regular spacetime, we can derive insights from the disformal transformation. Given that wormholes are characterized by non-homogeneous solutions, the disformal transformation offers a general framework for understanding the behavior of non-homogeneous spacetimes.
Using eq. (\ref{dismet}), we find that the construction of a wormhole in this case appears to be unattainable. This is because constructing a wormhole necessitates a finite and negative definite $g_{\tau\tau}$ as well as a minimum region at $r=r_0$. For this to be satisfied, it implies
\begin{equation}
\left(\frac{1}{1+2X W}\right)\bigg|_{r=r_0}=0\implies (1+2X W)\Big|_{r=r_0}\rightarrow\infty\implies W\Big|_{r=r_0}\rightarrow \infty
\end{equation}
 for finite $X$. Consequently, the $g_{\tau\tau}$ component blows up at the wormhole throat, rendering the construction impossible. 
 Possible ways to overcome this limitation include utilizing a seed $X$ that diverges at a finite double point while ensuring $W(X)$ remains finite everywhere. It is important to note that this implies the target 
 \begin{equation}
      X=\frac{X_{seed}}{1+2 W X_{seed}}
 \end{equation}
 remains regular throughout.
Another approach involves employing a seed metric with a $q$-dependent additional singularity, represented as $h(r)=h_{finite}-W(X)q^2$. In this case, the target metric remains well-behaved, and $W(X)$ can be employed to construct the wormhole throat.
Both of these approaches seem highly unlikely, suggesting a lack of viable wormhole solutions sourced from these configurations. If a seed solution capable of sourcing a wormhole exists, it is likely to possess highly pathological characteristics, like ill-defined action functionals at $X\rightarrow 0$.

\section{Breaking the Linear dependence}\label{nolinear}

Now that we have extracted the possible information from linearly dependent functionals, we are focusing on the more general problem. We shall refrain from presenting any actual solution, but rather focus on the strength of the algorithm presented, due to the high complexity of the results. We note that the equation (\Ref{eq2rev}) which provides the solution for $X$ is a generic algebraic equation. To keep the analysis tractable and driven by our motivation for exact compact-object solutions, we will consider configurations for $X$, when (\Ref{eq2rev}) yields a polynomial equation of order $m$ with respect to $X$. For this to be the case, we express $\Xi$ as a ratio of polynomials $\displaystyle \frac{P}{Q}$. Then, (\Ref{eq2rev}) is expressed as
\begin{equation}
\label{5.1}
    \frac{r^2 X  P(X)-q^2 Q(X)+2 X Q(X)}{Q(X)}=0\,.
\end{equation}
It is clear that in order for $X$ to be given via a polynomial equation of order $m$, two cases exist: Either $P$ and $Q$ are polynomials of order $m-1$ or $P$ and $Q$ are polynomials of order $m$ with $Q$ being singular at $X\rightarrow 0$, i.e. $Q$ does not contain  a constant term. The second case is problematic because it no longer yields a smooth GR limit when $q\rightarrow 0$. This is because $X$ will in general not vanish when $q=0$ and as such, fine tuning is required to recover GR solutions. For this reason, we drop this case and consider only the first one. Naturally, when (\Ref{5.1}) is a polynomial equation of the first order, the previous analysis is recovered. The logical extension here is to consider the cases when $X$ is given by a polynomial equation of second order under the constrain of $X\rightarrow0$ when $q\rightarrow0$. To this end, we set  

\begin{equation}
\label{5.2}
    \Xi=\frac{c X-a}{b X-d}
\end{equation}
for which the $X$ solution is
\begin{equation}
\label{5.3}
   X=\frac{a r^2+b q^2+2 d\pm\sqrt{\left(a r^2+b q^2+2 d\right)^2-4 d q^2 \left(2 b+c r^2\right)}}{4 b+2 c r^2}\,.
\end{equation}
Now, let $G_4$ be smoothly connected to GR for $X\rightarrow 0$ via
\begin{equation}
\label{5.4}
    G_4=\zeta+\sum_{n=1}^{\infty}g_{\frac{n}{s}}X^{\frac{n}{s}}\,,\hspace{1em} s\in \mathbb{Z}^+\,,
\end{equation}
in consistency with the previous analysis.
From the definition of the auxiliary function $\Xi$, (\Ref{jgzcond}), we extract a $G_2$, which reads
\begin{equation}
\label{5.5}
    G_2=-2\Lambda+\frac{a}{d}\sum_{n=1}^{\infty}g_{\frac{n}{s}}X^{\frac{n}{s}}+\frac{a b - c d}{d^2}\sum_{n=1}^{\infty}\frac{\frac{n}{s}}{\frac{n}{s}+1}g_{\frac{n}{s}}X^{\frac{n}{s}+1} \, _2F_1\left(1,\frac{n}{s}+1;\frac{n}{s}+2;\frac{b X}{d}\right)\,.
\end{equation}
Due to the existence of the hypergeometric function in $G_2$, either particular $\frac{n}{s}$ needs to be chosen in order to extract any possible exact configurations or one may fine-tune the parameters. We note that from the definition of $X$, if the coupling constant $c$ vanishes, $X$ blows up at infinity, while if $d$ vanishes, $X$ does not vanish for $q\rightarrow 0$.  On the other hand, if $b$ vanishes and $a$ does not, one is able to construct action configurations with a canonical kinetic term, although $X$ again does not vanish for $q\rightarrow 0$. Thus, one may verify that the choice of $b=0$, $a=0$ and $d=-\lambda^2$ yields a well behaved $X$ with a relatively manageable $G_2$ functional form of 
\begin{equation}
    \label{5.6}
    G_2=-2\Lambda+\frac{c}{\lambda^2}\sum_{n=1}^{\infty}g_{\frac{n}{s}}\frac{\frac{n}{s}}{\frac{n}{s}+1}X^{\frac{n}{s}+1}\,,
\end{equation}
while $X$ reads
\begin{equation}
\label{5.7}
    X=\frac{\sqrt{4 c \lambda ^2 q^2 r^2+4 \lambda ^4}-2 \lambda ^2}{2 c r^2}\,,
\end{equation}
where $c$ is positive and we kept the plus sign in the solution. Integrating (\Ref{eq3}), we found a complicated metric component given in terms of summations of hypergeometric functions that does not offer additional insight in being presented here. We conclude that, from the point of view of exact homogeneous local solutions, abandoning the linear dependence in the functionals yields highly complicated metric components, whose further analysis deviates from the scope of this article. The existence of square roots in $X$ makes any subsequent calculations increasingly unmanageable, due to the generality of our approach on the problem of local solutions in Beyond Horndeski gravity. However, we wish to pinpoint that the generality of the procedure allows one to in principle extract all possible static and spherically symmetric local solutions in the framework of shift and parity symmetric Beyond Horndeski gravity. Indeed, even under the constraint that we are working on a subclass agnostic framework, by considering a polynomial behavior of $\Xi(X)$, one can always find the $X$ configuration. Then, under the reasonable assumption that $G_4$ is smooth at $X\rightarrow 0$, we can extract the $G_2$ functional that supports such solutions. The final step is a relatively simple integration of (\Ref{eq3}) that yields the metric component.

\section{Conclusions}\label{conclusion}

 Our research has centered on exploring the comprehensive framework of shift-symmetric and parity-preserving Beyond Horndeski gravity, employing a local solution-generating algorithm. Within a semi-agnostic subclass of Beyond Horndeski gravity, characterized by a linear dependence of the $G_2$ and $G_4$ functionals and a smooth transition to General Relativity, we have successfully derived homogeneous static and spherically symmetric black hole solutions endowed with a primary charge. A notable finding is that these solutions can achieve regularity at a critical value of the black hole ADM mass, facilitated by the contribution of the primary charge. Furthermore, our investigations confirm that these configurations consistently adhere to the Weak Energy Conditions, establishing them as healthier compact objects in contrast to many hairy black holes within Horndeski gravity.

An essential outcome of our analysis is the revelation that in the examined generic subclass, a canonical kinetic term in the action generally leads to solid deficit angles. To circumvent such pathologies, we have demonstrated that a pure disformal transformation is adequate to produce inhomogeneous configurations with correct Minkowski asymptotics. This insight prompted a discussion on wormhole solutions within the broader framework of Beyond Horndeski gravity, originating from the primary charge. Indicative arguments were presented, suggesting the non-existence of such solutions in the generic framework when the seed action is well-defined and yields a proper General Relativity limit.

Finally, our exploration extended to generic $G_2$ and $G_4$ functionals, with a focus on theories yielding homogeneous solutions. We specifically constrained the kinetic term of the scalar field, $X=-\frac{1}{2}\partial_\mu\Phi\partial^\mu\Phi$ as a polynomial equation of the radial coordinate, ensuring a smooth transition to General Relativity vacuum solutions. Emphasizing the strength of the algorithm employed, our attention was directed towards the methodology rather than presenting specific local solutions, given the increasing complexity of the results.

In a compelling extension of our investigation, it would be intriguing to explore the shift and parity-breaking case within the realm of Beyond Horndeski gravity. An important facet of this exploration would involve testing whether the Weak Energy Conditions of black holes endowed with a primary charge can still be upheld in the presence of a Gauss-Bonnet term. Notably, the quadratic curvature nature of the Gauss-Bonnet term leads us to anticipate that the kinetic term of the scalar field will generally exhibit dependence on the metric components. This introduces a particularly intriguing expansion of the scalar charge to incorporate a correlation with the ADM mass.

Should this expectation materialize, it becomes highly plausible that wormholes featuring a dynamically varying throat contingent upon both the mass and the primary charge $q$ could manifest in this scenario, since the scalar configurations inherent in this context may be able to produce the necessary pathologies in the seed configurations in order to reach a wormhole solution in the disformed frame. Consequently, although wormholes with a dynamical throat seem impossible in the parity-preserving case, the potential existence of such configurations remains an intriguing prospect in the natural extension of the framework. It is important to underscore that a comprehensive analysis of these aspects is currently pending and represents a promising avenue for future exploration.

\section{Acknowledgements}
\noindent The research project was supported by the Hellenic Foundation for Research and Innovation (H.F.R.I.) under the “3rd Call for H.F.R.I. Research Projects to support Post-Doctoral Researchers” (Project Number: 7212). The work of N.C. is supported by the research project
of the National Technical University of Athens (NTUA) 65232600-ACT-MTG: {\it Alleviating Cosmological Tensions Through Modified Theories of Gravity}. We are very happy to thank 
Christos Charmousis and Panagiotis Dorlis for useful discussions. We also thank Alexandros Kehagias for pointing out a couple of unclear parts in our analysis.

\appendix

\section{Non-homogeneous solutions with primary scalar hair in shift-symmetric and parity-preserving Horndeski theories}\label{ap:horn-sols}

In this section, we are revisiting the non-homogeneous case in the context of shift-symmetric and parity-preserving Horndeski theory, hence, the theory that will occupy us is described by the action \eqref{eq:act} with $F_4(X)=0$. 
Also, we will constrain ourselves to theories with $\Xi=const.$, which by following the discussion in Sec. \ref{sec:gen-frame} means that
\begin{equation}
    \label{apA-eq:g4-g2}
    G_4 Z=\alpha+\lambda^2\,G_2 Z\,,
\end{equation}
where $\alpha$ is dimensionless and $[\lambda]=[L]$.
We recall that for the line element
\begin{equation}
    ds^2=-h(r)dt^2+\frac{dr^2}{f(r)}+r^2d\Omega^2
\end{equation}
the equations of motion are given by \eqref{eq1}-\eqref{eq3} with $Z=2X G_{4X}-G_4$. 
For this particular relation between functions $G_4$ and $G_2$, and for non-degenerate theories, eq.\,\eqref{eq2} can be rewritten as
\begin{equation}
    \label{apA-eq:eq2}
    Z^2 X=\frac{q^2\gamma^2}{2+(r/\lambda)^2}\,.
\end{equation}
The above relation forces to us the constraint $Z\neq c X^{-1/2}$.
Plugging this result into \eqref{eq3}, we obtain the very simple differential equation
\begin{equation}
    \label{apA-eq:eq3}
   2 \gamma ^2 \frac{\diff}{\diff r}\left[r h(r)\right]-\frac{\gamma^2  q^2}{X}-\alpha\frac{r^2}{\lambda^2}=0\,.
\end{equation}
 As it can be immediately verified, assuming homogeneous solutions in Horndeski, $X$ is linear in $q^2$ and, as such, no homogeneous black holes with primary charge can exist when $\Xi$ is constant. In particular, by the definition of the auxiliary $Z$ function, one may immediately verify that $Z$ is constant yields $G_4\sim \sqrt{X}$. Then, by (\Ref{apA-eq:g4-g2}), $G_2\sim \sqrt{X}$. This corresponds to a stealth Schwarzschild black hole, already presented in \cite{Bakopoulos:2023fmv}.
 Consequently, the simple set of equations \eqref{eq1}, \eqref{apA-eq:eq2}, and \eqref{apA-eq:eq3} can be used to obtain non-homogeneous compact-object solutions.


\bibliography{Refs}
\bibliographystyle{utphys}

\end{document}